# Droplet Impact Dynamics on Micropillared Hydrophobic Surfaces

Nagesh D. Patil, Rajneesh Bhardwaj*, Atul Sharma

Department of Mechanical Engineering,

Indian Institute of Technology Bombay, Mumbai, 400076 India

*Corresponding author (rajneesh.bhardwaj@iitb.ac.in)

Phone: +91 22 2576 7534, Fax: +91 22 2572 6875

***Abstract***

The effect of pitch of the pillars and impact velocity are studied for the impact dynamics of a microliter water droplet on a micropillared hydrophobic surface. The results are presented qualitatively by the high-speed photography and quantitatively by the temporal variation of wetted diameter and droplet height. A characterization of the transient quantitative results is a novel aspect of our work. Three distinct regimes, namely, non-bouncing, complete bouncing and partial bouncing are presented. A critical pitch as well as impact velocity exists for the transition from one regime to another. This is explained with a demonstration of Cassie to Wenzel wetting transition in which the liquid penetrates in the grooves between the pillars at larger pitch or impact velocity. The regimes are demarcated on a map of pitch and impact velocity. A good agreement is reported between the present measurements and published analytical models.

## 1 Introduction

Understanding impact dynamics of bouncing and non-bouncing droplets on hydrophobic and superhydrophobic surfaces is useful in several technical applications. These surfaces exhibit low wettability and this property may be leveraged in the applications such as pesticide spray coating [1], drag reduction [2], anti-snow adhesion surfaces [3], self-cleaning surfaces [4], and surface cooling via spray evaporative cooling [5] and on spatially varying wettability surfaces [6]. The fluid and interface dynamics during the droplet impingement on a solid surface is highly transient. For instance, impact of a 3 microliter isopropanol droplet with 0.37 m/s velocity takes around 7 ms to spread on a heated fused silica surface [7]. The liquid-gas interface exhibits capillary forces and dynamic wetting occurs at the contact line. During the impact, the droplet first spreads and the contact line recedes later, which may lead to the bouncing on surface with lower wettability. The advancing/outward motion is aided by the momentum or inertia-force during the spreading and is resisted by the surface tension and viscous forces. After the maximum spreading, the surface energy leads to the receding/inward motion of contact line and recoiling of the liquid-gas interface. During the change from advancing to receding motion of the contact line, the contact line is pinned to the surface and dynamic contact angle reduces from advancing to receding. Thus, the droplet fate depends on the interplay of several forces namely, inertia, viscous, surface tension and gravity, and surface wettability. The fate could be spreading/non-bouncing, partial bouncing, complete bouncing or splashing depending on the impact conditions and surface wettability (see review by Yarin [8]).

Previous experimental studies extensively investigated the droplet bouncing and non-bouncing on the hydrophobic as well as superhydrophobic surfaces. For instance, Richard and Quéré [9] recorded several bouncing cycles of a 1 mm water droplet on a superhydrophobic surface, with advancing contact angle of $170^o$. Similarly, Rioboo et al. [10] studied the effect of surface tension, viscosity and density on the surface with different wettabilities and found that the receding angle and surface roughness are two important parameters in deciding the bouncing. Renardy et al. [11] studied the bouncing of water droplets on superhydrophobic surfaces; and reported pyramidal as well as toroidal shapes of the droplet, after it bounces off at low and moderate impact velocities. They discussed internal swirling flow and thin film formation at the impact location. Clanet et al. [12] studied the bouncing of a 2.5 mm water droplet impacting with a velocity of 0.83 m/s on a superhydrophobic surface with equilibrium contact angle of $170^o$. They studied the maximum spreading of droplet for wide range of Weber numbers and proposed a model for the maximum width of the deforming droplet. Further, Antonini et al. [13] analyzed the droplet spreading and receding characteristics, during its impact on the hydrophilic to superhydrophobic surfaces for wider range of advancing contact angles (48-166°).

In the last decade, the droplet dynamics was investigated on engineered micropillared surfaces which exhibit varied surface wettability as a function of pillar diameter, pillar height and pitch. An important aspect is the wetting transition from Cassie [14] to Wenzel [15] state on such surfaces. The air gets trapped between the micropillars in the former while the liquid fills the complete region between the pillars in the latter. The characterization of the wetting transition was investigated in several studies without or with impact dynamics. For instance, the transition was recorded by increasing droplet volume [16], by evaporating sessile droplet [17-20] and, by varying the pillar pitch [18, 21] as well as pillar shape [22]. In the context of the reporting the effect of the wetting transition on the impact dynamics, previous studies showed that the wetting transition from Cassie to Wenzel state changes the droplet fate from bouncing to non-bouncing. For example, by varying the equilibrium contact angle of a carbon nanotube arrays [23], impact velocity [24-27, 28], pillar diameter [25, 29] and, pillar height and pitch [24, 25, 27].

The wetting transition described above was explained by first order analytical models. For example, Bartolo et al. [24] suggested that the transition occurs if dynamic pressure due to inertia ($p_d$) exceeds capillary pressure due to surface tension ($p_c$). Oftentimes, the partial bouncing was also recorded at larger impact velocities due to the partial penetration of the liquid in-between the pillars. Deng et al. [30] explained it using effective water hammer pressure ($p_{EWH}$) which generates due to the compression of droplet by shock wave at larger impact velocity. According to authors, the criteria for non-bouncing, complete bouncing and partial bouncing are $p_{EWH} > p_d > p_c$, $p_c > p_{EWH} > p_d$ and $p_{EWH} > p_c > p_d$, respectively. Further, Dash et al. [31] demonstrated a high capillary pressure for hollow square pillars in comparison with solid square pillars. They tested the stiffness of surfaces using a dynamic pressure and water hammer pressure, and showed that water hammer pressure depends on the surface morphology. Maitra et al. [32] showed that the fate of droplet partial bouncing or impalement on a pillared surface happened because of air compression (due to water hammer pressure) beneath the droplet. Very recently, Meng et al. [33] characterized the fate of droplet on three-level of hierarchical surfaces with varying pillar height. Their analysis showed that varying height and shape of micro-pillars can reduce the water hammer pressure and restricts the possibility of water penetration in-between the pillars.

As mentioned above, although there are numerous studies on the effect of surface morphology of the micropillared surfaces on the impact dynamics, there was almost no study on the effect of systematic variation of pillar pitch at various impact velocity. For example, pillar height, pillar diameter and pitch were varied simultaneously in Refs. [24, 25] and Tsai et al. [27] considered a smaller range of variation of pitch (1.8 to 7 µm) for a 5 µm square pillar. The present work considers the independent variation of pitch (30 to 76 µm) and impact velocity (0.22 to 0.62 m/s), at a constant cross-sectional dimension of 20 µm and height of 27 µm, for the square pillars. We discuss the three distinct droplet fates, namely, non-bouncing, complete bouncing and partial bouncing, measured for the above mentioned range.

To this end, the present experimental work has four objectives. First, perform a detailed qualitative as well as quantitative transient image (obtained from high speed visualization) analysis for various pitches of the pillars and the impact velocity; and study the role of surface wettability as well as kinetic energy and surface energy on the impact dynamics. Second, propose a regime map which demarcates various impact dynamics regimes in impact velocity-pitch plane. Third, demonstrate the droplet wetting transition from non-penetration to partial penetration to complete penetration of the droplet liquid in-between pillars; and analyze the corresponding outcome to non-bouncing/complete bouncing/partial bouncing. Fourth, propose a correlation for the maximum wetted diameter based on present measurements.

# 2 Experimental methods

## 2.1 Fabrication and characterization of micropillared surfaces

Micropillared surfaces are fabricated using ultraviolet lithography [34], after depositing SU-8 2025 epoxy photoresist polymer on polished side of 2 inch silicon wafer. This is done in a five-step process. First, Si wafer was cleaned by RCA cleaning process and wet oxidized in a furnace to remove suspended or dissolved components. Second, SU-8 was spin-coated on the wafer with a speed of 500 rpm for 10 s and 2300 rpm for next 40 s. Third, the wafer was soft-baked at 65°C for 3 min and at 95°C for next 8 min. An iron oxide coated glass mask with square-shaped patterns (printed using Laser Writer, LW405, Microtech Inc) was aligned on the top of spin-coated wafer using double sided aligner (EVG620, EV Group Inc). The wafer was exposed to UV radiation with intensity of 160 mJ/cm$^2$ for 2 to 3 min. Fourth, the samples were post-baked at 65°C for 1 min, 95°C for next 6 min and allowed to cool in ambient. Subsequently, they were developed in SU8 photo developer for 5 to 6 min and cleaned by isopropanol. Fifth, the wafer was kept on a heater at 120°C for 10 min for hard-baking and surfaces were coated with 10 nm platinum layer. Two different views of the fabricated surfaces – recorded by SEM – are shown in Fig. 1(a), for several pitches; pitch is the distance between centers of two adjacent pillars. The characterization of the pillars height and surface morphology on the micropillared surfaces is shown in Fig. 1(b) as 3D images as well as 2D line plot for cross-section surface profiles; the images were captured using 3D optical profilometer (Zeta-20, Zeta Instruments Inc.) for four different surfaces. The figure demonstrates that the surface profiles for each of the planes as well as all fabricated surfaces vary within a very narrow margin of the prescribed width and height of pillars. For all the surfaces, the measured width is 20±2 μm and height is 27±2 μm.

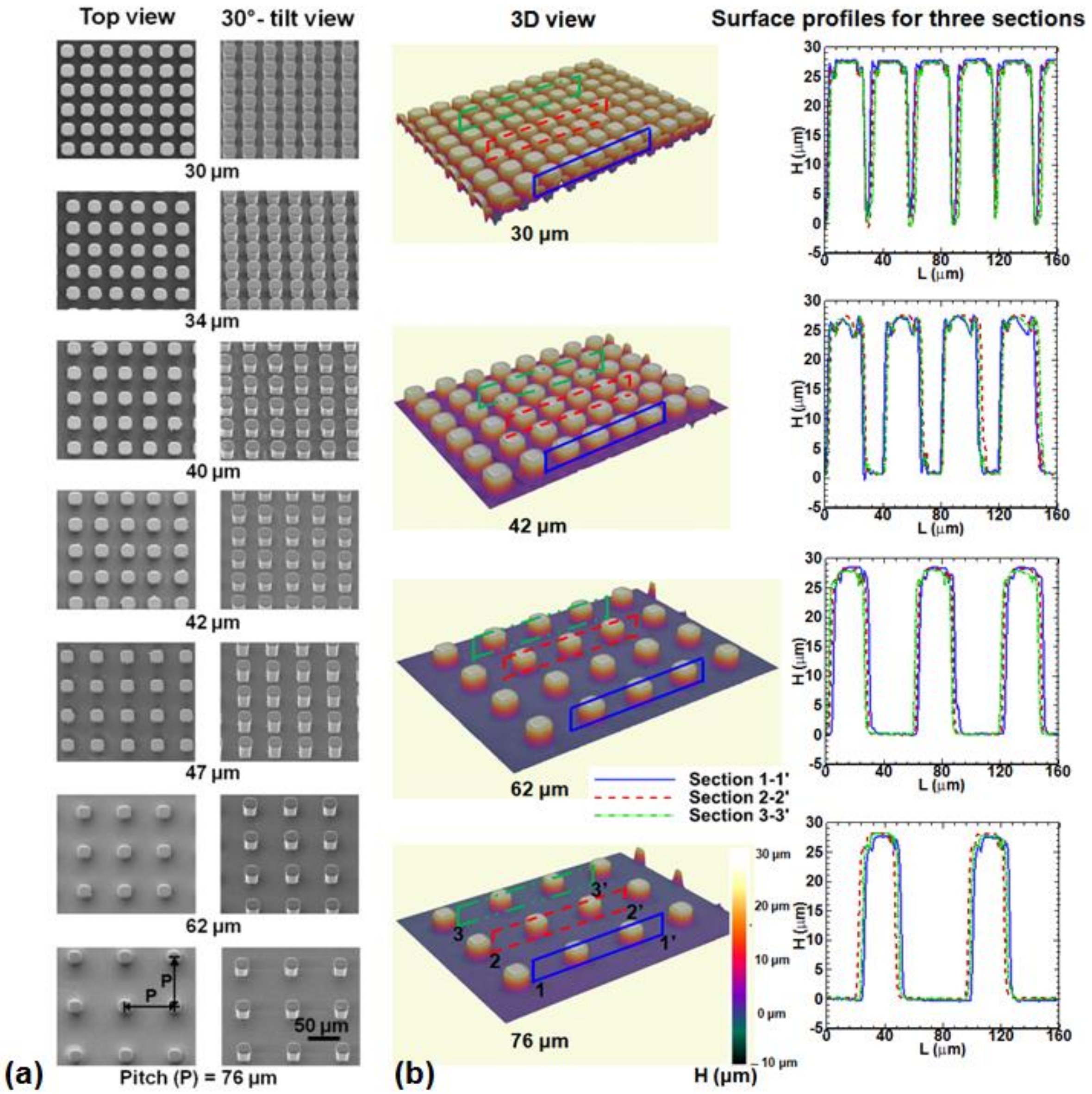

Fig. 1: (a) SEM and (b) 3D optical profilometer images of micropillared surfaces, for various pitches of uniformly-spaced square pillars (20 µm width and 27 µm height). Scale for SEM images is shown in the right bottom image of (a). Subfigure (b) shows 3D images (left) of micropillared surfaces (with pillars height topography) and corresponding 2D plot (right) for the surface profiles along three cross-sectional planes.

## 2.2 Droplet generation and high-speed visualization

Microliter deionized water droplets of 1.7±0.05 mm diameter are generated using a syringe, with 31 gauge needle. Adjusting the height of the syringe from the solid surface led to a variation of the impact velocity from 0.22 to 0.62 m/s, with an experimental uncertainty of ±0.01 m/s. The schematic of the experimental setup is shown in Fig. 2. The micropillared surfaces are carefully cleaned with isopropanol and allowed to dry out completely. The droplets were visualized from the side using a high-speed camera (MotionPro, Y-3 classic, CMOS, C-mount) with long distance working objective (Qioptiq Inc.); similar to setup in Refs. [35, 36]. A white LED lamp serves as a back light source. The selected magnification corresponds to 14 μm per pixel, which implies dimensional error of ±28 μm. The images of 192 × 632 pixels at 1500 frames per second are recorded with exposure time of 330 μs.

In order to study the effect of different pitches for the outcome of hydrophobicity, equilibrium contact angles ($\theta_{\mathrm{eq}}$) on the fabricated surfaces are measured after a gently deposited droplet on the surface became a sessile spherical cap [37, 38]; as shown in Fig. 3a. Note that the flat surface is the processed silicon wafer with 10 nm platinum layer coating, as described in section 2.1. Further, to determine the contact angle hysteresis ($\theta_H = \theta_{adv} - \theta_{rec}$) of the surface, the advancing ($\theta_{adv}$) and receding ($\theta_{rec}$) contact angles are measured by tilting the surface with droplet kept on it; as done by Jung and Bhushan [26]. A microscopic evaluation of the contact angles is done, looking at the tangent of the contact line as shown in Fig. 3b. The obtained contact angle values are plotted in Fig. 3c. The figure shows $\theta_{\mathrm{eq}}$= 95° for a flat surface (0 μm pitch), which increases from around 139 to 150° with an increase in the pitch from 30 to 76 μm for the micropillared surfaces. Whereas, the contact angle hysteresis ($\theta_H$) decreases from 41 to 21° with an increases in the pitch, indicating that the surface wettability decreases due to the presence of micropillars. The uncertainty in the contact angle measurements is ±3°. All measurements are performed three times to ensure repeatability in a controlled environment at 25 ± 1°C and 38 ± 5% relative humidity.

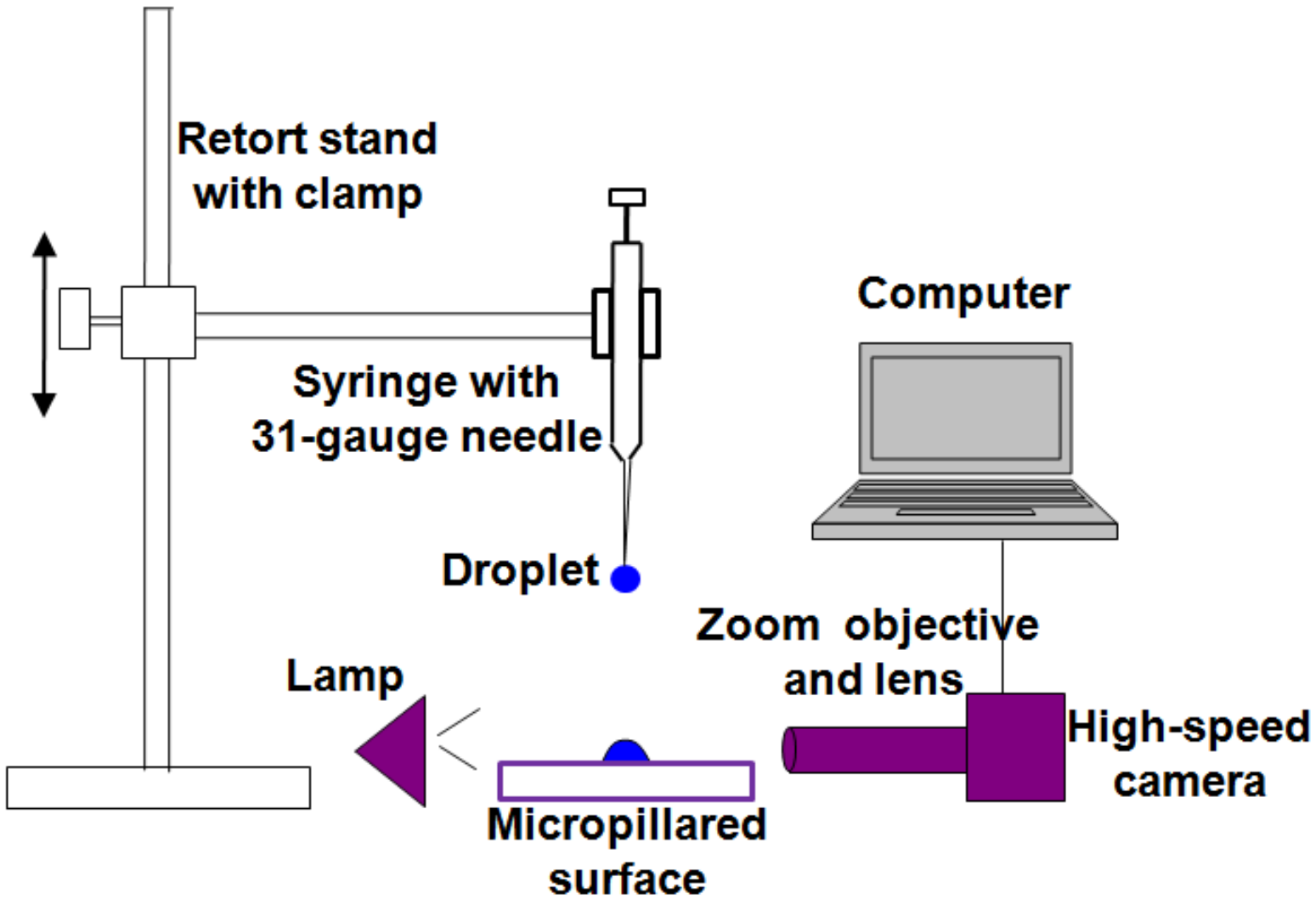

Fig. 2: Schematic of the present experimental setup.

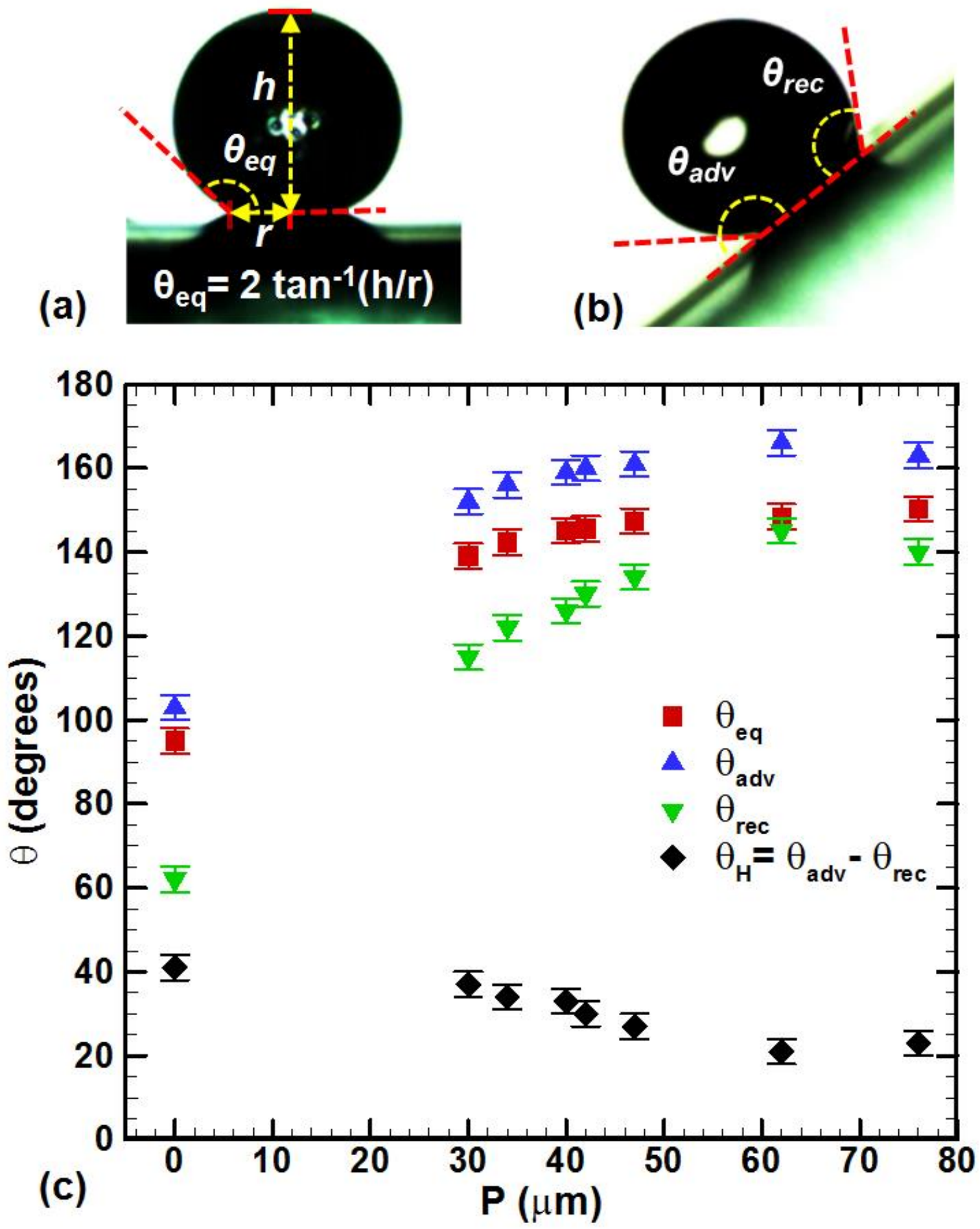

Fig. 3: (a) Sessile droplet on a micropillared surface for the measurement of equilibrium contact angle; (b) sliding droplet on an inclined stage for the measurement of advancing and receding contact angles; and (c) measured equilibrium, advancing and receding contact angles for flat surface and micropillared surfaces (with several pitches *P*). Contact angle hysteresis is also shown in Fig. (c). The uncertainty in the measurement of *P* and *θ* are ±2 μm and ±3°, respectively.

# 3 Results

In this section, results for the impact of microliter water droplet of initial diameter 1.7 mm on flat and micropillared surfaces are presented for various values of the governing parameters:

*Pitches*: 0 (flat surface), 30, 34, 40, 42, 47, 62 and 76 µm

*Impact velocities*: 0.22, 0.34, 0.43, 0.52 and 0.62 m/s.

For the above dimensional parametric range, the variation of the non-dimensional parameters is 377-1065 for the Reynolds number ($Re = \rho U_0 D_0 \mu^{-1}$) and 1.2-9.3 for the Weber number ($We = \rho U_0^2 D_0 \gamma^{-1}$). Here, $\rho$, $U_0$, $D_0$, $\mu$ and $\gamma$ are density, impact velocity, initial droplet diameter, dynamic viscosity and surface tension, respectively. The Ohnesorge number, $Oh = \mu(\rho\gamma D_0)^{-0.5}$ is 0.002858 for all the measurements reported in the present work. The effect of the pitch as well as impact velocity on the droplet dynamics is presented below in separate subsections. The experimental results are also compared with the published analytical models.

### 3.1 Effect of the pitch of the pillars

Figure 4(a) presents qualitative comparison among the droplet shapes recorded using high speed visualization on flat as well as micropillared surfaces with several pitches at an impact velocity of 0.34 m/s (see also supplementary data[1]). A horizontal green line demarcates the droplet from its reflection on the surface as shown in the figure. The droplet spreads on all the surfaces until $t$ = 3.33 ms and the maximum spreading is almost same on the micropillared surfaces. The flat surface exhibits 27% larger spreading as compared to micropillared ones due to its larger wettability ($\theta_{\text{adv}}$ = 103°, Fig. 3c) as compared to the latter ($\theta_{\text{adv}}$ = 152-166°, Fig. 3c).

As shown in Fig. 4(a), the contact line recedes from $t$ = 4 to 8 ms and after 8 ms, the droplet bounces off on 40, 47 and 62 µm surfaces while it does not bounce on flat, 30 and 76 µm surfaces. The role of wettability on the contact line motion changes from aiding for the spreading to opposing for the receding. Thus, lesser wettable (larger $\theta_{adv}$) micropillared surface offers lesser opposition to the receding as compared to the flat surface. The droplet bounces off on all micropillared surfaces except on ones with the smallest and the largest pitch. At the smallest pitch of 30 µm, the decrease in the wettability as compared to that for the surfaces with intermediate pitches is not sufficient for the bouncing. Whereas, at the largest pitch of 76 µm, the droplet liquid penetrates in grooves in-between the pillars and the droplet is unable to pull or break-off the penetrated liquid. In other words, the Cassie to Wenzel wetting transition occurs at 76 um pitch, leading to the non-bouncing; demonstrated and discussed later in section 4.2.

---

[1] Electronic supplementary information (ESI) available: Movie 1 - High-speed visualizations of impact of a 1.7 mm water droplet with 0.34 m/s impact velocity on flat and micropillared surfaces with several pitches.

In Fig. 4(b), the quantitative analysis of the visualized images for the impact dynamics is presented by plotting temporal variation of the non-dimensional wetted diameter $D_{wetted,}$ (with respect to initial droplet diameter; marked in the last column of Fig. 4(a) at $t$ = 3.33 ms) for all surfaces. The instantaneous droplet shapes are shown for eight time-instances in Fig. 4(a) while the symbols in the respective line plots are shown for 46 time-instances in Fig. 4(b). Note that the time duration considered in Fig. 4(b) is much larger as compared to Fig. 4(a), in order to analyze the dynamics of the bouncing droplet after the first impact. Furthermore, the results for the micropillared surfaces with pitches of 34 and 42 µm are also included in Fig. 4(b). Note that the zero value of $D_{wetted}$ corresponds to the bouncing and plateau value of $D_{wetted}$ after a certain time indicates that the droplet has become sessile on the surface.

As compared to micropillared surface, the duration of the spreading as well as receding on the flat surface is almost same; however, the maximum (minimum) wetted diameter at the end of the spreading (receding) is much larger for the flat (micropillared) surface (Fig. 4(b)). The minimum wetted diameter decreases to 1.17 on the flat, 0.38 on 30 µm surface and 0.29 on 34 µm surface, due to descending wettability (as $\theta_{rec}$ increases from 62 to 122°, respectively, Fig. 3c). The time-instance at the onset of first bouncing is close to 9.33 ms - shown as the onset of zero wetted diameters - for 40, 42, 47 and 62 µm surfaces in Fig. 4(b); the time decreases slightly with increasing pitch. Thereafter, the wetted diameter remains zero until certain time-duration and latter becomes non-zero - indicating second impact of the droplet. This time-duration corresponds to a vertically upward followed by downward (due to gravity) movement of the droplet. It can be seen that the duration increases substantially with the increasing pitch; the time instant for the second impact is 27.33 ms on 40 µm, 29.33 ms on 42 µm, 39.33 ms on 47 µm and 46.67 ms on 62 µm surfaces. The early second impact at smaller pitch as compared to larger pitch is due to larger $\theta_H$ (Fig. 3c) which results in larger wettability at the beginning of the first bouncing (which occurs almost at the same time instances for all the surfaces). After the second impact, Fig. 4(b) shows that the droplet does not re-bounce, except for the surface with a pitch of 62 µm.

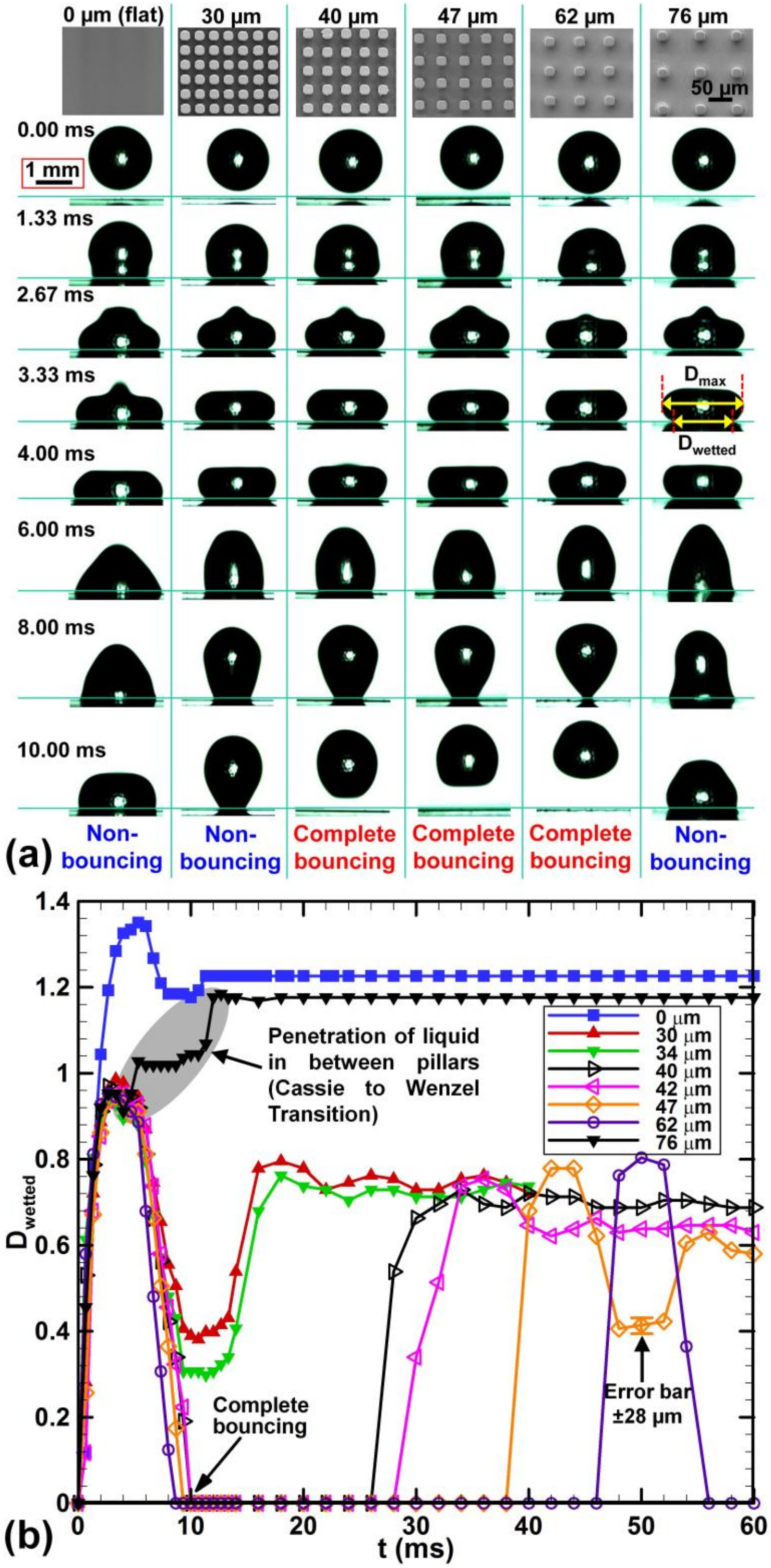

Fig. 4: (a) High-speed visualization of the instantaneous interface, during impact of a 1.7 mm water droplet (with a velocity of 0.34 m/s), on a flat surface and micropillared surfaces (of various pitches). Scale is shown in the left top image. (b) Temporal variation of dimensionless wetted diameter for all cases. Filled and hollow symbols represent non-bouncing and complete bouncing of the droplet, respectively.

As shown by a shaded region in Fig. 4(b), temporal variation of $D_{wetted}$ for 76 µm pitch behaves substantially different and it indicates the Cassie to Wenzel transition, responsible for non-bouncing. The time-variation of $D_{wetted}$ on this surface shows almost monotonic increase except for a slight decrease around $t$ = 3.33 ms and this indicates a much smaller receding as compared to that on all other surfaces. With further increase in time, the figure shows a spreading rate (obtained by the ratio of change in wetted diameter and the time duration during the spreading) which is quite large at the beginning and at the end of the impact dynamics; and is almost zero with no-spreading for quite some time duration (indicated by almost constant $D_{wetted}$ = 1.02 from $t$ = 6 to 8.67 ms). The extreme slowdown of the spreading rate is conjectured here to correspond to the large time-duration needed for the penetration of the droplet fluid – in-between the pillars - with almost no increase in the wetted diameter.

Various quantitative parameters obtained from the present experiments are found to be in good agreement with available analytical models in the literature. The measured values of non-dimensional maximum droplet width $D_{\max}$ at 3.33 ms in Fig. 4(a) are around 1.4 for flat and 1.26 for micropillared surfaces, which are in good agreement with the model ($D_{\max} \sim We^{1/4}$) of Clanet et al. [12], $D_{\max,\ \mathrm{analytical}}$ ~ 1.29. The measured time-periods of the free surface oscillation for non-bouncing cases are around 8.67 ms and contact times of the bouncing are 9.33 ms (Fig. 4(a) and 4(b)). These measured times are on the same order as those given by the models in the literature [39, 40] ($t \sim \sqrt{\rho D_0^3 / \gamma} = 8.26$ ms).

### 3.2 Effect of the impact velocity

The effect of impact velocity is investigated for five cases - 0.22, 0.34, 0.43, 0.52 and 0.62 m/s. The time-varying droplet shapes and wetted diameter are compared in Fig. 5(a) and 5(b), respectively. The symbols in the Fig. 5(b) correspond to the data extracted from 45 images, obtained from high-speed visualization[2]. After the first impact, Fig. 5(a) shows non-bouncing, complete bouncing and partial bouncing at small, intermediate and large impact velocity, respectively. At 4 ms, the droplet assumes an ellipsoidal shape at smallest velocity (0.22 m/s) and toroidal/puddle at larger velocities (0.34-0.62 m/s); similar to the published results [12, 27].

[2] Electronic supplementary information (ESI) available: Movie 2 - High-speed visualizations of impact of a 1.7 mm water droplet on a micropillared surface with 42 micrometer pitch with several impact velocities.

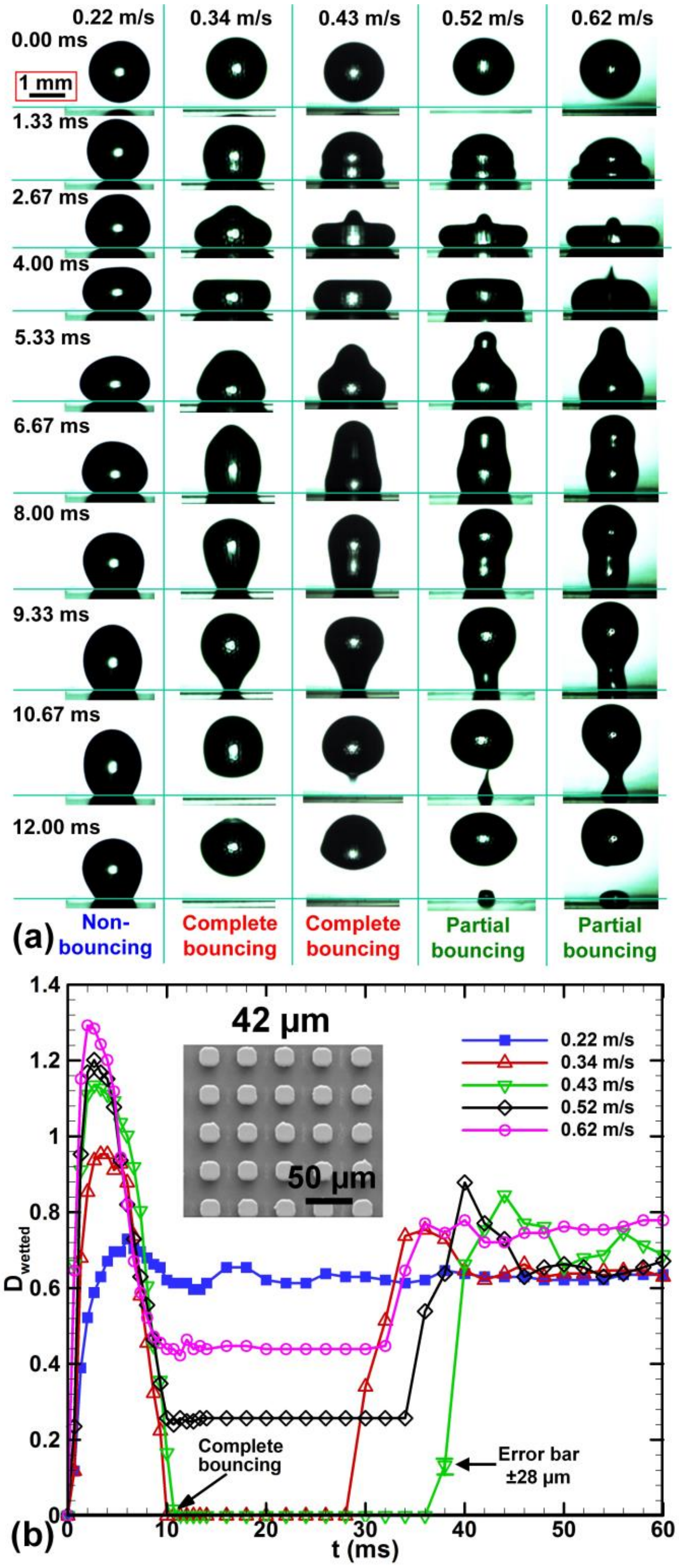

Fig. 5: (a) High-speed visualization of the instantaneous interface, during impact of a 1.7 mm water droplet on a micropillared surface (with 42 µm pitch) at various impact velocities. Scale is shown in the left top image. (b) Temporal variation of dimensionless wetted diameter for all cases. Filled and hollow symbols represent non-bouncing and complete/partial bouncing of droplet, respectively.

After the first impact, a monotonic increase of the wetted diameter in Fig. 5(b) corresponds to the spreading and the monotonic decrease thereafter corresponds to the receding. At the end of the spreading, the maximum wetted diameter increases and the corresponding time taken for the maximum spreading decreases with increasing impact velocity. For instance, the time taken for 0.22 m/s and 0.62 m/s cases are 6.0 and 2.67 ms, respectively. However, it can be seen that the receding ends at almost same time-instance (~10 ms) for all cases of impact velocities. Thus, the time taken for the receding increases with increasing impact velocity. Furthermore, during receding, it can be seen that the change in the wetted diameter increases with an increase in impact velocity up to 0.43 m/s and decreases with further increase in the impact velocity.

This leads to an increase in the average receding rate (obtained by the ratio of change in wetted diameter and the time duration during the receding) as 0.16, 0.15, 0.12, and 0.10 at 0.34, 0.43, 0.52 and 0.62 m/s, respectively. Thus, the average receding rate is larger in the complete bouncing (at 0.34 and 0.43 m/s) and smaller in the partial bouncing regimes (at 0.52 and 0.62 m/s). The slowdown of the receding rate in the partial bouncing as compared to the complete bouncing regime is due to an additional time required to pull or break-off the droplet liquid penetrated in-between the pillars.

After receding, bouncing is indicated in Fig. 5(b) by a constant value of $D_{wetted}$ for some time duration; zero for complete bouncing and non-zero for partial bouncing at intermediate and larger values of the impact velocity, respectively. The zero and non-zero value indicate non-penetration and partial penetration of the droplet liquid in-between the pillars, respectively. Partial penetration is defined as the Cassie to Wenzel wetting transition - corresponding to the liquid filling the region between only the central (not the circumferential) pillars underneath the droplet. During receding, the penetrated liquid gets pinned to the surface and it leads to break-up of the droplet into a primary droplet which moves upward (bounces) and a secondary droplet remains on the surface (Fig. 5(a) for 0.52 and 0.62 m/s).

The influence of the impact velocity on the droplet dynamics can be explained in terms of associated energies [41-43]. Before the droplet impact, the total energy of the droplet consists of kinetic energy and surface energy. Thereafter, the impact induced spreading motion of the droplet decelerates with ~ $U_0^2/D_0$ [12]; and at the instance of maximum spreading, the kinetic energy of the droplet becomes negligible as it gets transferred to the surface energy [12, 40, 41].

At this quasi-static stage (where the droplet is about to recede), according to the energy conservation principle, the sum of total energy of the droplet just before the impact becomes almost equal to the sum of surface energy and energy lost due to the viscous dissipation [41, 43]. Thus, the surface energy of the droplet basically depends on the initial kinetic energy (i.e. impact velocity) and the solid surface energy (i.e. surface wettability). In Fig. 5(a), for the constant droplet diameter and impact on the same surface, the surface energy is expected to increase with an increase in the impact velocity from 0.22 to 0.34 m/s or 0.43 m/s, which leads to the transition from the non-bouncing to the complete bouncing. With further increase in impact velocity from 0.52 to 0.62 m/s, the increased initial kinetic energy leads to a change from no-penetration to the partial penetration and thereby shows partial bouncing (discussed later in section 4.2). Thus, with an increase in impact velocity, the surface energy of the bouncing droplets increases for the complete bouncing and decreases for the partial bouncing. The decrease of surface energy for the partial bouncing regime may be due to the smaller size of the bouncing droplet as well as energy spent in the break-up of the liquid-gas interface.

For the partial bouncing regime, Fig. 5(b) shows that the wetted diameter of the secondary droplet increases (thus, the diameter of the primary droplet decreases), with increasing impact velocity. The duration for the upward followed by downward movement of the bouncing droplet is represented by the duration for the constant value of $D_{wetted}$ in Fig. 5(b); it increases for the complete bouncing and decreases for the partial bouncing, with increasing impact velocity. In Fig. 5(b), after the constant value of $D_{wetted}$, its increasing trend of variation indicates spreading after the second impact. The increase is followed by an oscillatory trend of variation of $D_{wetted}$, before it asymptotes to a plateau value. Note that the plateau value of $D_{wetted}$ is almost same for all cases of the impact velocity; except a much larger value at the largest velocity, which corresponds to partial bouncing.

The various quantitative parameters are compared with those obtained by the published analytical models, shown in Fig. 6. The figure shows good agreement between the present measurement and the model proposed by Clanet et al. [12], for the increase in $D_{\max}$ with increasing impact velocity; with a maximum error of 13%. It can also be seen that the time-period of the free surface oscillation for the non-bouncing case ($t_{osc}$) and the time-instant at the onset of complete bouncing ($t_b$) (~ 10 ms in Fig. 5(b)) are slightly larger than that predicted by the analytical model [39, 40] ($t \sim \sqrt{\rho D_0^3 / \gamma} = 8.26\ \text{ms}$).

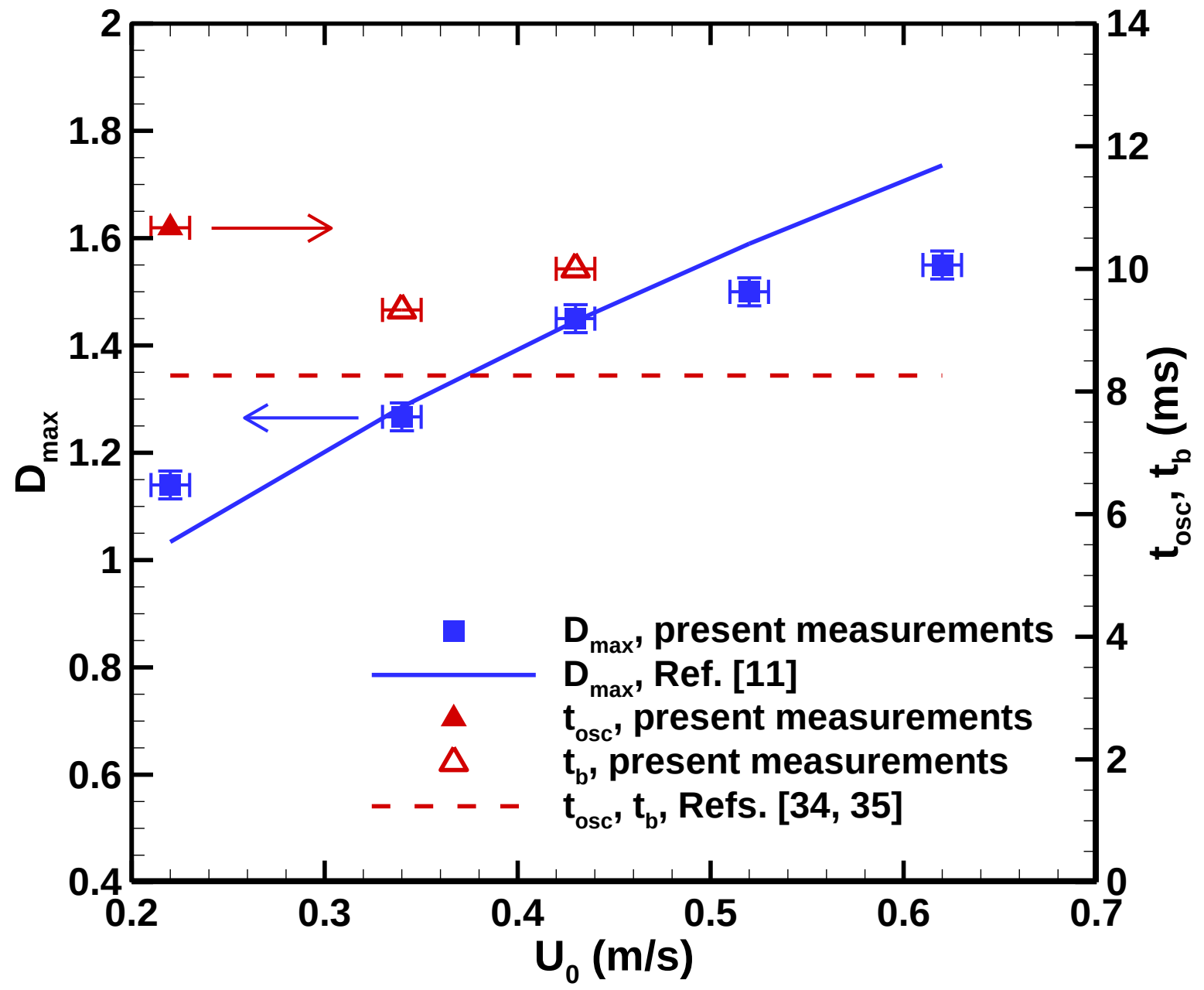

Fig. 6: Comparison of present measurements with published analytical models, for variation of maximum dimensionless droplet width ($D_{max}$), time-instant at the onset of complete bouncing ($t_b$) and time-period of the free surface oscillation for the non-bouncing case ($t_{osc}$), with increasing impact velocity ($U_0$) on a micropillared surface with 42 µm pitch. The error bars represent the uncertainty in the measurement of $D_{max}$ and $U_0$ are ±28 µm and ±0.01 m/s, respectively.

## 3.3 Regime map

In this section, a regime map is proposed in Fig. 7 with data of 40 experiments for several impact velocities (0.22 to 0.62 m/s) and pitches of the pillars (0 to 76 μm). Fig. 7 shows the demarcation of three distinct regimes, namely, non-bouncing, complete bouncing and partial bouncing. The map shows the transition pattern to the regimes as follows,

Non-bouncing → Complete bouncing → Partial bouncing → Non-bouncing

With increasing pitch, the map shows the transition from one regime to the other at a certain critical value of pitch, which decreases with increasing impact velocity. In addition, with increasing impact velocity, the map shows the transition at a critical impact velocity, which also decreases with increasing pitch. This variation in the critical values of the pitch and impact velocity are due to increase in the droplet kinetic energy, with increasing impact velocity and increase in penetrability of liquid in the pillars, with increasing pitch, respectively.

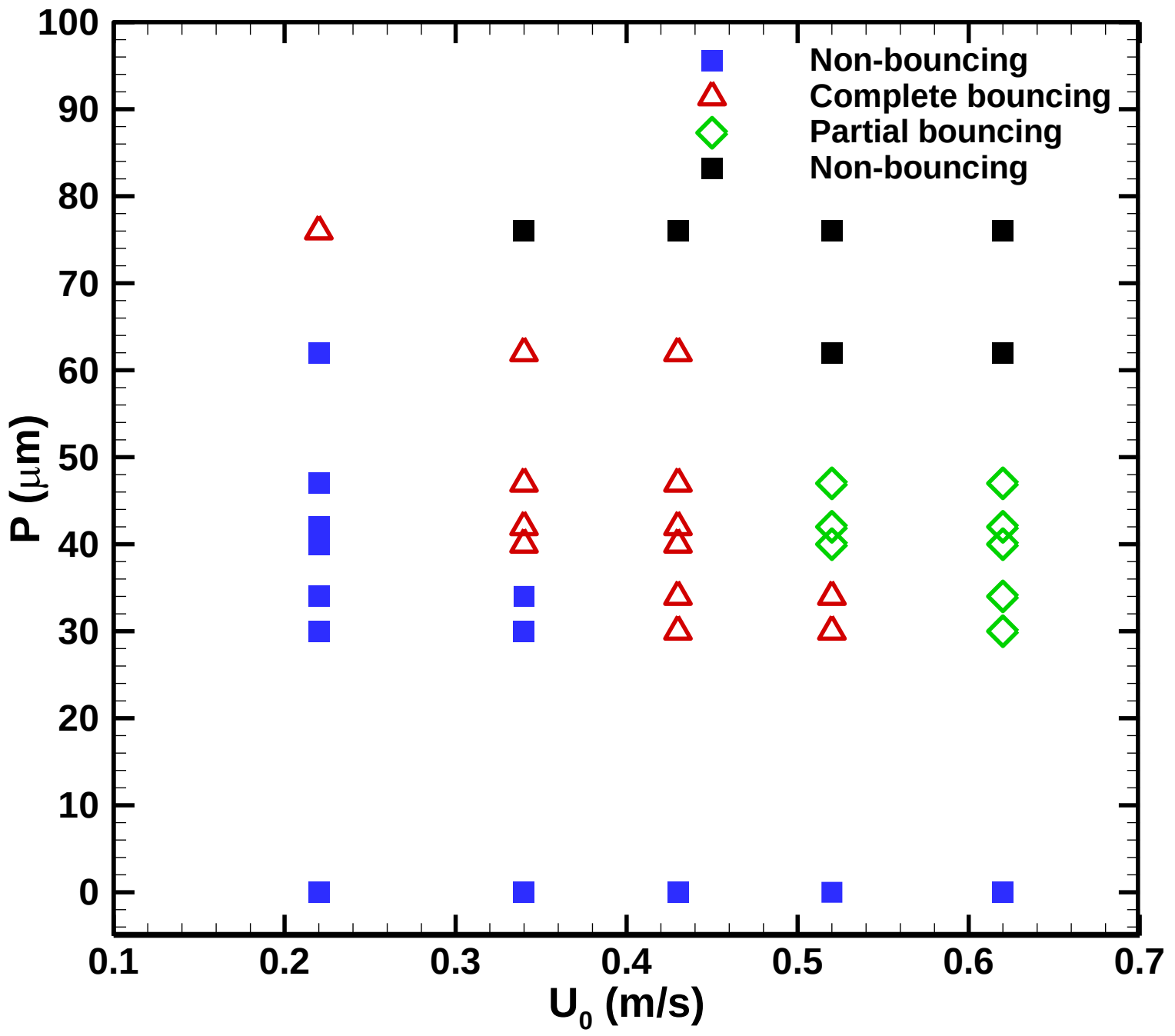

Fig. 7: Droplet impact dynamics regime map for impact of 1.7 mm water droplet on the flat and micropillared surfaces. The uncertainty in the measurement of $P$ and $U_0$ are ±2 µm and ±0.01 m/s, respectively.

# 4 Discussion

## 4.1 Analysis of non-bouncing, complete bouncing and partial bouncing

An analysis is presented with the help of a time-wise synchronized variation of the qualitative visualized droplet shapes in Fig. 8(a-j) as well as the quantitative non-dimensional wetted-diameter and the droplet maximum height in Fig. 8(k). Fig. 8(a-j) shows the comparison among representative cases of the non-bouncing, complete bouncing and partial bouncing corresponding to 0 (flat), 30 and 47 µm pitch surface, respectively. The impact velocity is kept constant at 0.52 m/s for all cases in Fig. 8. In Fig. 8(k), the temporal variation of the wetted-diameter as well as the droplet maximum height is compared for these three cases. Note that the line in the figure corresponds to the present experimental results with a temporal resolution of 0.67 ms; whereas, the symbols correspond to the time instances at which the visualized images are shown in Fig. 8(a-j).

The droplet spreading is shown in Fig. 8(a-c) until $t$ = 2.67 ms, with a rapid increase in the wetted diameter and decrease in the droplet height; shown quantitatively in Fig. 8($k_1$-$k_3$).

However, at the end of the spreading, Fig. 8(c) shows a larger wetted diameter and almost same droplet height for flat ($D_{wetted}$ ~ 1.47 and $H_{max}$ ~ 0.50, from Fig. 8($k_1$)) as compared to micropillared ($D_{wetted}$ ~ 1.2, and $H_{max}$ ~ 0.53, from Fig. 8($k_2$-$k_3$) surface. This is due to the larger wettability (lower $\theta_{eq}$ and higher $\theta_H$, Fig. 3c) of the flat surface; thus, the spreading phenomenon is attributed to not only the kinetic energy of droplet but also the surface wettability.

During the receding, the surface energy dominates and the contact line recedes in all the cases (Fig. 8(c-e)). The receding is noted with a rapid decrease in the wetted diameter and an increase in the droplet height – in Fig. 8($k_1$-$k_3$); in-between the time duration corresponding to the symbols "c-e". It is interesting to note a much smaller decrease in the wetted diameter for the time duration between "c" and "d" as compared to that between "d" and "e" for the complete bouncing (Fig. 8($k_2$)); vice-versa for the partial bouncing (Fig. 8($k_3$)). The substantial slowdown on the receding rate - for the partial as compared to complete bouncing case - is due to the additional time required to pull and break-off from the partially-penetrated fluid. Since the role of wettability changes from aiding for spreading to opposing for receding, the larger wettable flat surface shows a smaller change in wetted diameter during receding motion (Fig. 8(k)).

Figure 8(e-f) shows an early bounce off for the complete bouncing as compared to the partial bouncing case. Whereas, for the non-bouncing case, Fig. 8($k_1$) shows that the droplet free surface oscillates with fixed wetted diameter ($D_{wetted}$ ~ 1.2). Note that on the 30 µm surface, the droplet is in Cassie state during the spreading as well as the receding stage resulting in complete bouncing. Whereas, the droplet on the 47 µm surface is subjected to the partial penetration during the impact itself (due to larger kinetic energy and larger pitch) and the penetrated liquid remains stuck during the receding; leaving a smaller chunk of liquid at the impact location and remaining mass of the droplet bounce off from the surface (Fig. 8($a_3$-$g_3$)).

Finally, on both the micropillared surfaces (30 and 47 µm), Fig. 8($k_2$-$k_3$) indicates that the bouncing droplet returns back to the surface and the droplet free surface oscillates till it attains steady state – with an almost same wetted diameter $D_{wetted}$ ~ 0.75 in the both cases. The amplitude of vertical oscillation shown by a periodic variation of $H_{max}$ in Fig. 8($k_1$-$k_3$) dissipates with time. However, after the second impact at "$g_2$" for the complete bouncing and "$g_3$" for the partial bouncing, the amplitude of the periodic variation of $H_{max}$ is substantially reduced for the partial bouncing as compared to the complete bouncing – with an almost constant value of 0.83 for both the cases. The reduction in the amplitude is attributed to dissipation in the kinetic energy during the partial penetration in the partial bouncing case. However, for the complete bouncing case after the first impact, competition of the kinetic and surface energy results in oscillating behavior of $H_{max}$.

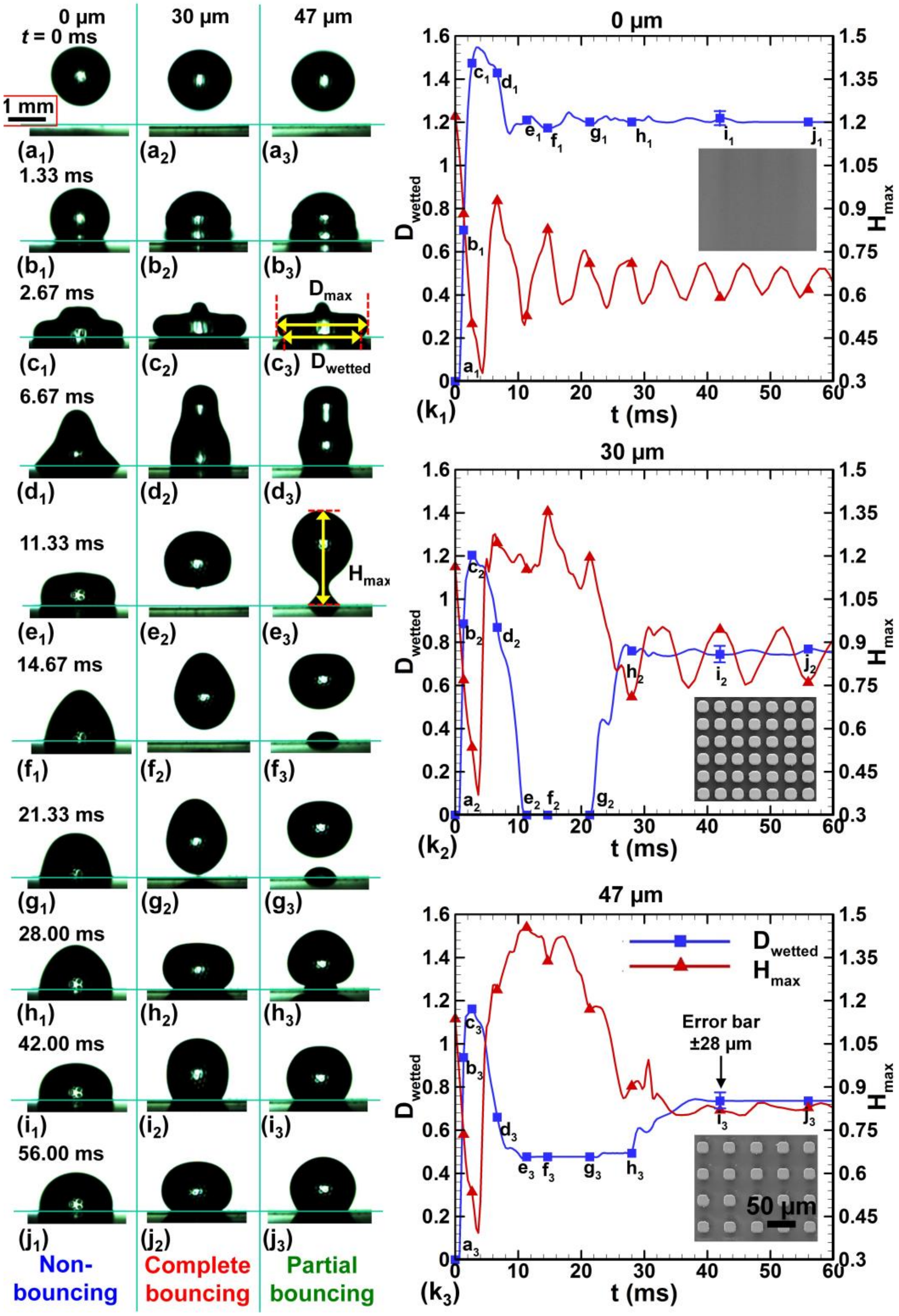

Fig. 8: (a-j) High-speed visualization of the instantaneous interface, during impact of a 1.7 mm water droplet with a velocity of 0.52 m/s on a flat surface ($a_1$-$j_1$) and micropillared surfaces ($a_2$-$j_2$ and $a_3$-$j_3$). Sub-figures ($k_1$-$k_3$) represents temporal variation of dimensionless wetted diameter and maximum height of droplet, with the symbols corresponding to the time instant for the instantaneous interface plots.

Hence, for a particular impact velocity, an increase in the pitch of the pillars increases the possibility of the complete bouncing; however, after a critical value of pitch, the droplet may first lead to the partial penetration and subsequently at larger pitch to the complete penetration leading to the partial bouncing and non-bouncing, respectively. For instance, further increase in the pitch from 47 μm to 76 μm in Fig. 7 - for the impact velocity of 0.52 m/s – results in the transition from partial to complete penetration and the non-bouncing of the droplet.

## 4.2 Transition from non- to partial- to complete-penetration of the droplet in the grooves between the pillars

As discussed above, the droplet fate depends on the interaction of the liquid-gas interface with the grooves between the pillars of the micropillared surface. The interaction depends on the impact velocity as well as the pitch, leading to three possible dynamic states of the droplet, shown schematically in Fig. 9(a).

In Fig. 9(a), the states are classified based on the penetration of the liquid in the grooves, namely as Cassie, partial wetting and Wenzel state corresponding to non-penetration, partial penetration and complete penetration, respectively. The grooves are completely filled by the entrapped air for the first and by the liquid for the third state. In the second state, the grooves are filled by the liquid in the central region and by the air in circumferential region. In addition, the completely air-filled grooves in the first state and partially air-filled grooves in the second state lead to complete and partial bouncing, respectively. The completely liquid-filled grooves in the third state lead to non-bouncing. The transition from non- to partial-/complete- penetration occurs if the dynamic pressure exceeds the capillary pressure across the liquid-gas interface [24], which results in collapsing of the interface inside grooves between adjoining micropillars (Fig. 9(b)).

An evidence of the schematic in Fig 9(a) is presented for a representative case of Cassie to Wenzel transition in Fig. 9(c) and Cassie to partial wetting transition in Fig. 9(d). The figures show actual and zoomed-in view of the images obtained using high-speed visualization. Fig. 9(c) shows time-wise penetration of the droplet liquid in the grooves for 0.34 m/s impact velocity on 76 µm surface, which demonstrates the transition from non-penetration/Cassie to complete penetration/Wenzel state. The zoomed-in view in the insets shows that most of the grooves are filled by air from $t$ = 2.67 to 3.33 ms and by the liquid at $t$ = 4 and 4.67 ms. In this context, the wetting transition was demonstrated by Jung and Bhushan [18] using similar visualization technique for a water droplet on a patterned silicon surface. Thus, first the liquid gradually penetrates in between the central pillars and subsequently in circumferential pillars. The penetration occurs during the spreading and is due to dominance of the kinetic energy over the surface energy which holds the liquid-gas interface on the pillars. Finally, during receding, the droplet is unable to pull or break-off the penetrated liquid and shows non-bouncing.

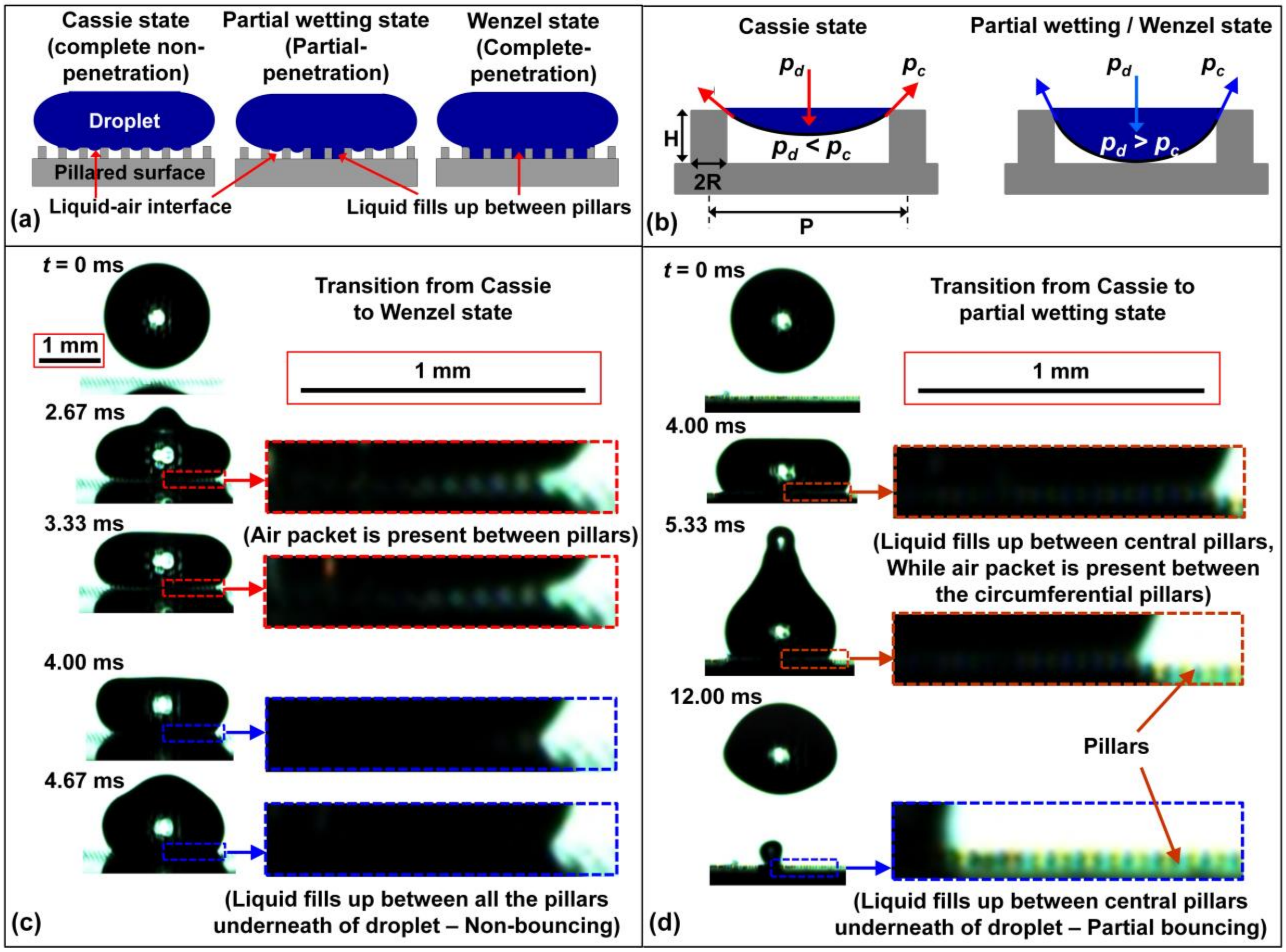

Fig. 9: Schematic for (a) three different states of impact dynamics of a droplet on a micropillared surface; and (b) droplet wetting transition based on a model by Bartolo et al. [24]. High-speed visualization of the transition for 1.7 mm water droplet impacting with a velocity of (c) 0.34 m/s on a 76 µm pitch and (d) 0.52 m/s on 40 µm pitch micropillared surface.

The Cassie to partial wetting transition is shown in Fig. 9(d), for 0.52 m/s impact velocity on 40 µm surface. During spreading, the kinetic energy of the droplet is sufficient enough to penetrate the liquid in the central region but not in the circumferential region as shown in the zoomed-in view of in Fig. 9(d) at $t$ = 4 ms. Thereafter, the receding starts and the penetrated liquid does not have sufficient energy to pull off from the surface. It remains stuck in between the pillars while remaining mass of droplet bounce off from the surface, leaving a small amount of the liquid on the surface; as shown in zoomed-in view of Fig. 9(d) at $t$ = 12.00 ms.

In Fig. 9(b), an analytical model proposed by Bartolo et al. [24] is utilized to confirm the Cassie to Wenzel wetting transition on larger pitch surface (76 µm), According to the model, if dynamic pressure, $p_d \sim 0.5\rho U_0^2$ exceeds, capillary pressure, $p_c \sim \gamma HR / P^3$, the transition occurs from the complete bouncing/Cassie to non-bouncing/Wenzel, where $H$, $R$ and $P$ are height, radius and pitch of the pillars, respectively. For the 76 µm surface, the computed values of $p_d$ are 24, 58, 92, 135 and 192 Pa, for impact velocities of 0.22, 0.34, 0.43, 0.52 and 0.62 m/s, respectively; and $p_c$ is 38 Pa. Since $p_d < p_c$ for 0.22 m/s and $p_d > p_c$ for other impact velocities, the model predicts complete bouncing (Cassie state, Fig. 9(b), left) and non-bouncing (Wenzel state, Fig. 9(b), right), respectively. Our measurements presented in Fig. 7 match well with predictions by the model.

### 4.3 Correlation between the maximum wetted diameter, Weber number and quasi-static contact angle

In this section, a correlation between the maximum wetted diameter, Weber number and a quasi-static contact angle is presented. Such correlation is useful for estimation of wetted area on the surface and relevant to various technical applications; namely, ink-jet printing, self-cleaning, cooling of hot surfaces, etc. During the droplet spreading on the micropillared surface, in the stage of maximum deformation, the droplet shape becomes flattened and looks similar to a puddle [12] as shown in Fig. 10(a). Clanet et al. [12] presented an analytical model to estimate maximum droplet width as $D_{max} \sim D_0 We^{1/4}$. In Fig. 10(b), the comparisons of variation of $D_{max}$ predicted by Clanet et al. [12] model (solid black line) and obtained by measurements (filled symbols) are presented for all micropillared surfaces. The comparison is good for all measurements with a maximum error of 13% at the largest Weber number.

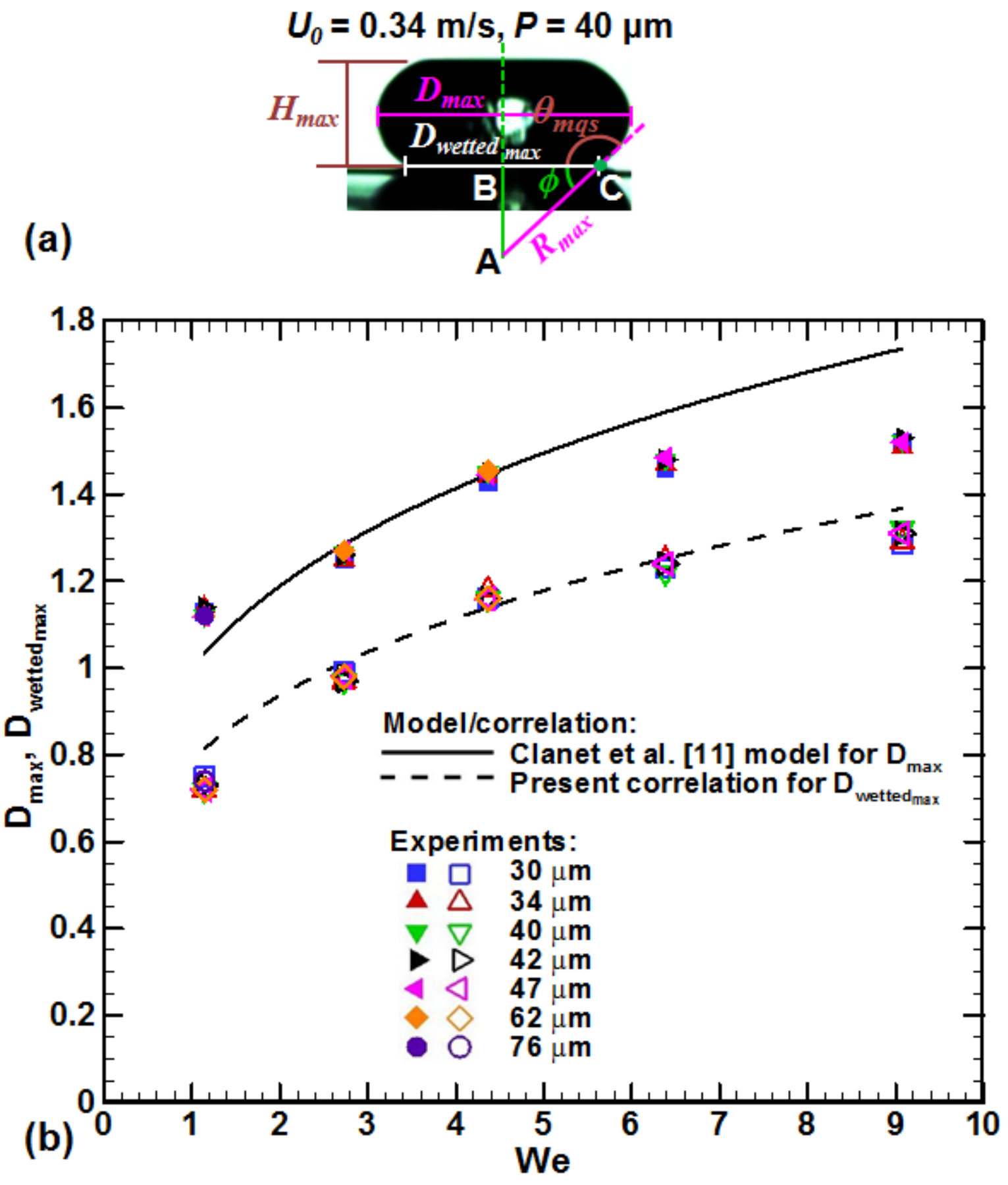

Fig. 10: (a) Schematic of the variables considered in the present image analysis, at the instant of maximum spreading/wetting. (b) Variation of maximum dimensionless droplet width ($D_{max}$) and dimensionless wetted diameter ($D_{wetted_{max}}$), with increasing Weber number for various micropillared surfaces. Present experimental results are compared with Clanet et al. [12] model and with present correlation. The filled symbols represent $D_{max}$ and the hollow symbols represent $D_{wetted_{max}}$.

The key to our proposition of a correlation is our finding – from the image analysis on all surfaces - that the base and hypotenuse of the triangle ABC (Fig. 10(a)) are almost equal to the maximum wetted radius ($R_{wetted_{max}} \sim 0.5 D_{wetted_{max}}$) and the maximum droplet radius ($R_{max} \sim 0.5D_{max}$), respectively. The angle $\phi$ at the contact line is measured as $\phi = 180 - \theta_{mqs}$, where $\theta_{mqs}$ is a quasi-static contact angle at the maximum spreading where droplet is passing from spreading to receding stage. The distance $R_{max}$ corresponds to a tangent drawn at the contact point C. From the present image analysis and trigonometric relations, $D_{wetted_{max}}$ can be calculated as follows:

$$D_{wetted_{max}} \approx -D_{\max} \cos\theta_{mqs} \approx D_{\max} \left|\cos\theta_{mqs}\right| \tag{1}$$

Substituting $D_{max} \sim D_0 We^{1/4}$ from the scaling analysis of Clanet et al. [12], the above equation is given as

$$D_{wetted_{\max}} \approx D_0 We^{1/4} \left| cos\theta_{mqs} \right| \tag{2}$$

This correlation is plotted in Fig. 10(b) (dashed black line), along with the present experimental results of $D_{wetted_{\max}}$ (hollow symbols). In our experiments, the measured $\theta_{mqs}$ for all the micropillared surfaces at the instance of maximum spreading lies within a narrow range of 139 to 146°. Thus, the mean value of 142.5° was considered while plotting the Eq. (2) in Fig. 10(b). It can be seen that the experimental results for all the surfaces (30-76 µm) collapse to a single curve. Note that the results for 62 and 76 µm surfaces are not shown in the figure for larger *We*, as more wetting is expected due to the transition from Cassie to Wenzel state. Fig. 10(b) shows a good agreement between the present correlation and our experimental results, with a maximum error of 10% at lowest Weber number. This confirms that the maximum wetted diameter ($D_{wetted_{\max}}$) for the impact of a water droplet - on a higher contact angle surfaces (139 to 150°) - scales as $D_0 We^{1/4}|cos\theta_{mqs}|$.

## 5 Conclusions

The impact dynamics of a microliter water droplet - with 1.7 mm initial diameter - on flat as well as micropillared hydrophobic surfaces is investigated experimentally. The surfaces are fabricated using ultraviolet lithography with pitch of the pillars and equilibrium contact angle in range of [0 - 76 µm] and [139 - 150°], respectively. In the measurements, the range of the impact velocity is [0.22 - 0.62 m/s]. The impact dynamics is studied qualitatively by high-speed photography as well as quantitatively by characterizing the time-wise variation of the droplet dimensions during spreading, receding, bouncing, second impact and rebouncing. In particular, the effect of the pitch of the pillars as well as impact velocity is investigated systematically in order to understand the interplay of surface wettability with kinetic and surface energy. The conclusions drawn from our study are as follows.

1. The following regimes are obtained with increasing pitch at constant velocity or with increasing impact velocity at constant pitch,

   Non-bouncing → Complete bouncing → Partial bouncing → Non-bouncing

2. The smaller as well as larger pitch leads to non-bouncing regime and intermediate pitch leads to the bouncing regime. Similarly, non-bouncing is recorded at smaller and larger impact velocity, while the droplet bounces off the surface at intermediate impact velocity.
3. The above regimes are demarcated in a regime map on impact velocity - pitch plane, demonstrating the existence of a critical pitch as well as impact velocity for the transition from one regime to the other. The critical pitch (impact velocity) for the transition from one to the next regime decreases with increasing velocity (pitch).
4. At larger impact velocity and pitch, the Cassie to Wenzel wetting transition occurs in which the liquid penetrates in the grooves between the pillars. The complete and partial penetration of the liquid inside the grooves corresponds to the non-bouncing and partial bouncing, respectively. During the transition, the wetted diameter corresponds to slowdown of the droplet spreading due to the complete penetration for the non-bouncing. Similarly, the slowdown of the receding of the contact line occurs due to the partial penetration for the partial bouncing.
5. Various quantitative parameters obtained from the experiments, namely, the droplet maximum width, bouncing time, time-period of the free surface oscillation, are in good agreement with the existing analytical models. A correlation between the maximum wetted diameter, Weber number and a quasi-static contact angle is proposed and found to be suitable in the parameter range of present measurements.

## 6 Acknowledgements

R.B. gratefully acknowledges financial support from Department of Science and Technology (DST), New Delhi through fast track scheme for young scientists. N.D.P. was supported by a Ph.D. fellowship awarded by Industrial Research and Consultancy Centre, IIT Bombay. The micropillared surfaces were fabricated at Centre of Excellence in Nanoelectronics and SEM images were taken at Centre for Research in Nanotechnology and Science, IIT Bombay. We thank Prof. Ramesh K. Singh, IIT Bombay for providing access to optical profilometer in his laboratory, as well as Mr. P. G. Bange for assistance in the experiments.

## 8 Nomenclature

**Symbols**

| | |
|---|---|
| $D_0$ | initial droplet diameter [m] |
| $D_{max}$ | non-dimensional maximum droplet width/diameter |
| $D_{wetted}$ | non-dimensional droplet wetted diameter |
| $D_{wetted_{\max}}$ | non-dimensional maximum droplet wetted diameter |
| $H_{max}$ | non-dimensional maximum axi-symmetric droplet height |
| $Oh$ | Ohnesorge number |
| $p_d$ | dynamic pressure [Pa] |
| $p_c$ | capillary pressure [Pa] |
| $p_{EWH}$ | effective water hammer pressure [Pa] |
| $P$ | pitch of the pillars [μm] |
| $Re$ | Reynolds number |
| $t$ | time [s] |
| $t_{osc}$ | time-period of the free surface oscillation for non-bouncing droplet [s] |
| $t_b$ | contact time for bouncing droplet [s] |
| $U_0$ | impact velocity of a droplet [m/s] |
| $We$ | Weber number |

**Greek letters**

| | |
|---|---|
| $\gamma$ | surface tension [N/m] |
| $\theta_{adv}$ | advancing contact angle [°] |
| $\theta_{eq}$ | equilibrium contact angle [°] |
| $\theta_{mqs}$ | quasi-static contact angle at the maximum spreading [°] |
| $\theta_{rec}$ | receding contact angle [°] |
| $\theta_H$ | contact angle hysteresis [°] |
| $\mu$ | dynamic viscosity [Pa.s] |
| $\rho$ | density [kg/m$^3$] |

[1] Bergeron et al. (2000)
[2] Truesdell et al. (2006)
[3] Antonini et al. (2011)
[4] Bixler and Bhushan (2014)
[5] Patil and Bhardwaj (2014)
[6] Attinger et al. (2014)
[7] Bhardwaj et al. (2010)
[8] Yarin (2006)
[9] Richard and Quéré (2000)
[10] Rioboo et al. (2001)
[11] Renardy et al. (2003)
[12] Clanet et al. (2004)
[13] Antonini et al. (2012)
[14] Cassie and Baxter (1944)
[15] Wenzel (1936)
[16] Lafuma and Quéré (2003)
[17] Callies and Quéré (2005)
[18] Jung and Bhushan (2007)
[19] Anantharaju et al. (2009)
[20] Xu et al. (2013)
[21] Lee et al. (2010)
[22] Zheng et al. (2005)
[23] Wang et al. (2007)
[24] Bartolo et al. (2006)
[25] Jung and Bhushan (2008)
[26] Jung and Bhushan (2009)
[27] Tsai et al. (2009)
[28] Nosonovsky and Bhushan (2008)
[29] Rayssat et al. (2006)
[30] Deng et al. (2009)
[31] Dash et al. (2012)
[32] Maitra et al. (2014)
[33] Meng et al. (2014)
[34] Milionis et al. (2013)
[35] Bhardwaj et al. (2009)
[36] Bhardwaj et al. (2010)
[37] De Gennes et al. (2003)
[38] Bhardwaj et al. (2008)
[39] Rayleigh (1879)
[40] Richard et al. (2002)
[41] Pasandideh-Fard et al. (1996)
[42] Ukiwe and Kwok (2005)
[43] Lee and Lee (2011)